\documentstyle[prb,aps,manuscript,twocolumn]{revtex}  
\begin{document} 
\title{Magnetic and electrical resistance behaviour of the oxides, Ca$_{3-x}$Y$_x$LiRuO$_6$ (x= 0.0, 0.5 and 1.0)}
\author{M. Mahesh Kumar and E.V. Sampathkumaran}
\address{Tata Institute of Fundamental Research,  Homi  Bhabha  Road,
Mumbai-400005, INDIA}
\maketitle
\tighten
\begin{abstract}
\noindent{We have investigated the magnetic and electrical resistance behaviour of  Ca$_{3-x}$Y$_x$LiRuO$_6$. The parent compound exhibits magnetic ordering from Ru sublattice at a rather high temperature, 113 K. Though the paramagnetic Curie temperature ($\theta$$_p$) is negative indicative of antiferromagnetic ordering, the large magnitude (-250 K) of $\theta$$_p$  reveals complex nature of the magnetism in this compound. Ru ions appear to be in the pentavalent state.  We note that the N\'eel temperature undergoes only a marginal reduction by Y substitution. All these compositions are found to be insulators and thus the electron doping does not result in metallicity. Thus the overall  magnetic and transport behaviour are found to be essentially insensitive to Y substitution for Ca, a finding which may favour the idea of quasi-one-dimensional magnetism in these compounds.}      
\end{abstract}
 
\section{Introduction}
There is a considerable interest in the current literature in identifying Ru-based magnetic and/or superconducting oxides. The magnetism in a Ru-based material for the first time was reported nearly forty years ago, namely, in SrRuO$_3$ exhibiting ferromagnetism below (T$_C$=) 160 K [1]. Subsequently there has been very little work in this direction until the discovery [2] of superconductivity at (T$_c$=)  1.5 K in Sr$_2$RuO$_4$, which turned out to be the first superconductor not containing Cu with the same structure as by-now-well-known K$_2$NiF$_4$-related high-T$_c$ oxides; this work naturally rekindled the interest on Ru-based oxides. The investigations in recent years led to the discovery of the following Ru based materials with the magnetism arising from Ru: Sr$_4$Ru$_3$O$_{10}$, T$_C$= 148 K [3]; Sr$_3$Ru$_2$O$_7$, T$_C$= 104 K [4], Eu(Gd)Sr$_2$RuCu$_2$O$_8$, T$_N$= 168 (185)K [5], YSr$_2$RuO$_6$, T$_N$= 26 K [7], and A$_3$A'RuO$_6$ (A= Ca, Sr; A'= Li, Na), T$_N$= 70-120 K [8]. While the ferromagnets of this class  are generally metallic, the antiferromagnets are found to be insulators. However, there is no distinct evidence for high-T$_c$ superconductivity (from Ru d band) in any Ru-based oxides. At this point, it is worth noting that superconductivity could be induced  in the temperature (T) range 30-50 K by small Cu doping in the antiferromagnetic insulator, YSr$_2$RuO$_6$, and in fact superconductivity and magnetism seem to compete in such substituted oxides [7]; however, considering that Cu doping is essential for superconductivity, one is not sure whether this phenomenon arises from Ru. Also, the compounds, R$_{1.4}$Ce$_{0.6}$RuSr$_2$Cu$_2$O$_{10}$ (R= Eu and Gd) are found to be superconducting below about 40 K  arising from Cu layers, whereas Ru exhibits magnetic ordering below about 180 K [6]. Thus, one may conclude that there is no clearcut evidence for high-T$_c$ in Ru oxides not containing Cu and that there is an upsurge of activity in identifying novel Ru based oxides.  In this context, we report here the results of magnetization (M) and electrical resistivity ($\rho$) on a compound, Ca$_3$LiRuO$_4$ [8], hitherto not extensively studied in the literature; we have also probed the influence of partial Y substitution for Ca on its properties in order to explore whether metallicity (and possibly superconductivity) could be induced by electron doping, considering that such doping effects have profound influnce on the properties of many other oxides, e.g., by now well-known, cuprates and manganates. 

	The compound under investigation has been synthesized and reported to adopt K$_4$CdCl$_6$-type rhombohedral structure [8]. The crystallographic details can be found in Ref. 7. The structure consists of infinite chains of alternating face-sharing LiO$_6$ trigonal prisms and RuO$_6$ octahedra (antitrigonal prisms), which are separated by Ca ions. 
\section{Experimental}
	The compounds have been prepared by solid state method as described by Darriet et al [8]. To start with, CaRuO$_3$ was prepared by heating in air of appropriate amounts of high purity CaCO$_3$ (Leico Industries, 99.995\%) and RuO$_2$ (Cerac, 99.9\%) at 750 $^o$C for 24 hours and then for 8 days at 1100 $^o$C in air. The samples were then prepared by reaction of CaRuO$_3$, Li$_2$CO$_3$ (Koch Chemicals, 99.999 \%) and Y$_2$O$_3$ (Johnson-Matthey, 99.9\%) at 550 $^o$C for 24 hours, at 800 $^o$C for 24 hours and at 950 $^o$C for two weeks with intermediate grindings. All the preparations were carried out under a flow of oxygen. The samples were subsequently characterized by x-ray diffraction (Cu K$_\alpha$) and scanning electron microscope. The magnetization measurements were performed by a commercial (Quantum Design) Superconducting Quantum Interference Device (SQUID) Magnetometer in the T interval 5 - 300 K. We have obtained the magnetization data for all the compositions in different ways (isothermal magnetization (M), T-dependence of magnetic susceptibility ($\chi$) at different magnetic fields (H), field-cooled (FC) and zero-field-cooled (ZFC) $\chi$ behaviour) and we show only those data which are relevant to highlight the main findings. The temperature (T= 77-300 K)) dependent electrical resistivity ($\rho$) behaviour was probed  by a conventional four-probe method, employing silver paint to make electrical contacts. 
\section{Results and Discussion}
The results of x-ray diffraction measurements are shown in Fig. 1. The diffraction pattern for the parent compound could be indexed on the basis of K$_4$CdCl$_6$-type structure except perhaps for the presence of few very weak unidentified lines ($<$5\%); the diffraction pattern seen by us is identical to the one reported by Darriet et al [8] and the lattice constants (a= 9.221 \AA  and c= 10.798 \AA) are also in agreement with their report. A piece of the parent compound was subjected to the 950 $^o$C (24 hrs) heat treatment in air and  brought to room temperature by quenching and this heat treatment apparently resulted in purer sample, as indicated by the disappearance of the weak extra line (marked by asterisk in Fig. 1).  The Y substituted samples also are characterized by the same x-ray diffraction pattern as that of that of the parent compound, though few, weak additional lines start appearing in the range 2$\Theta$ = 30 to 40 degrees, the origin of which is at present unclear. The x-dependence of unit-cell volume clearly reveals that there is a contraction of the unit-cell (along c-direction, see Fig. 1) with Y substitution, expected for replacement of (bigger) Ca ions.       The homogeniety of the samples were further checked by scanning electron spectroscopy and we do not find evidence for segregation of any other phase. The energy dispersive x-ray analysis in addition established that the proportion of the metallic elements are in good agreement with the starting compositions and uniform throughout the sample. We therefore conclude that the properties reported here are characteristic of the pure phases. 
 
	In figure 2a we show the T dependent $\chi$ behavior recorded in the presence of a field of 2 kOe for the ZFC as well as FC state of a specimen of the parent compound. It is distinctly clear that there is a sharp rise at 113 K as T is lowered with a peak at about 107 K for the ZFC state, however without any such peak for FC state; thus ZFC-FC curves deviate below 113 K in perfect agreement with the experimental observations of Darriet et al [8]. This observation establishes that we have prepared a sample with the same magnetic characteristics  as those of these authors. Thus, these data provide evidence for the fact that there is a magnetic ordering setting in below (T$_N$=) 113 K in this compound. In order to see how T$_N$ is influenced by Y substitution, we have recorded the FC data very carefully in a field of 100 Oe and the data (see Fig. 2b) suggest that the value of T$_N$ undergoes only a marginal reduction with increasing x (108 and 106 K for x= 0.5 and 1.0 respectively). With respect to $\chi$ behaviour in the paramagnetic state, the plot of inverse $\chi$ versus T is found to be linear in the range 160-300 K (see Fig. 3) and the value of the effective moment ($\mu$$_{eff}$) and paramagnetic Curie temperature ($\theta$$_p$)  obtained from the region 150-300 K are nearly independent of x (3.96, 3.65 and 3.77 $\mu$$_B$ and -250, -180 and -207 K for x= 0.0, 0.5 and 1.0 respectively).  It may be noted that the value of $\mu$$_{eff}$ is indicative of pentavalent state of Ru ion, assuming spin-only contribution to magnetic moment. It is important to note that the sign of $\theta$$_p$ is negative, which suggests that the exchange interaction is of an antiferromagnetic type. However, the value of T$_N$ is far below the magnitude of $\theta$$_p$, which may indicate complex nature of the magnetism of this compound. In order to get better insight on the nature of magnetic ordering, we have also performed isothermal M measurements at various temperatures for the ZFC state of the samples at many temperatures and typical behaviour are shown in Fig. 4. It is obvious that, at 120 K, M varies linearly with H as expected for a paramagnetic state; for T$<$100 K, M is a non-linear function of H for initial applications of H (below about 20 kOe), however undergoing linear variation for higher values of H, without any indication for saturation. This behaviour of M establishes that these compounds are better classified as antiferromagnets, though we observe some hysteresis loops  as shown in the inset of Fig. 4, for instance for x= 0.0 and 0.5.  

	Thus, all these results establish that the magnetism of Ru ions is fairly insensitive to Y substitution at the Ca site. Thus magnetic Ru chains are decoupled by intervening Ca ions. This finding  may support the idea of quasi-one-dimensional behaviour of magnetism in this compound, a question of debate in this class of compounds [8].

	We now present the results of $\rho$ measurements (Fig. 5). It is obvious that the Y substituted compounds remain insulating like the parent compound, exhibiting activated behaviour with the activation energy marginally decreasing by about 10 meV by Y substitution (for both x= 0.5 and 1.0) from 90 meV for x= 0.0. Interestingly, the $\rho$ of quenched specimen of the parent compound  is significantly lower and the activation energy is also considerably reduced to 1.5 meV (Fig. 6).   

\section{conclusions}
	To conclude, the magnetic and electrical transport behaviour of the oxides  Ca$_{3-x}$Y$_x$LiRuO$_6$ (x= 0.0, 0.5 and 1.0) have been investigated. These are one of the very few Ru-based oxides exhibiting magnetic ordering at rather high temperatures, presumably of a complex type. The carrier doping induced by Y substitution for Ca apparently does not bring about any significant change in the magnetic and transport behaviour of the parent compound. On the basis of the present results, we  infer that the Ru-magnetism could be of quasi-one-dimensional character.     
Finally, we would like to point out an observation in the low-field magnetization data for the ZFC-state of the specimen (see Fig. 4): In the magnetically ordered state, below about 400 Oe, the sign of M is negative (which is however positive for the FC specimen). Though a possible source  of this negative M could be that the specimen may not be in true ZFC state due to a small negative remanent field in the SQUID magnetometer we employed, this view can be challenged by the observation that the data at 120 K (in the paramagnetic state) do not show negative M at low fields. Alternatively, the magnetic moments (below T$_N$) in the ZFC state  may have got locked in the direction of the negative remanence field requiring larger applications of H in the positive direction to reorient them. More work is required to understand this aspect of these compounds.  

\newpage
\section{References}	                
\begin{enumerate}

\item{J.T. Randall and R. Ward, J. Am. Chem. Soc., {\bf81} (1959) 2629.}

\item {Y. Maeno, H. Hashimoto, I.K. Yoshida, S. Ishizaki, T. Fujita, J.G. Bednorz, and F. Lichtenberg, Nature (London), {\bf372} (1994) 532.}

\item {G. Cao, S.K. McCall, J.E. Crow, and R.P. Guertin, Phys. Rev. B, {\bf56} (1997) R5740.}

\item{G. Cao, S. McCall, and J.E. Crow, Phys. Rev. B, {\bf55} (1997) R672.}

\item{I. Felner, U. Asaf, S. Reich, and Y. Tsabba, Physica C,{\bf311} (1999) 163.}

\item{I. Felner, U. Asaf, Y. Levi, and O. Millo, Phys. Rev. B, {\bf55} (1997) R3374.}

\item{M.F. Wu, D.Y. Chen, F.Z. Chien, S.R. Sheen, D.C. Ling, C.Y. Tai, G.Y. Tseng, D.H. Chen, and F.C. Zhang, Z. Phys. B, {\bf102} (1997) 37.}

\item{J. Darriet, F. Grasset, and P.D. Battle, Mater. Res. Bull., {\bf32} (1997) 139.}

\end{enumerate}

\begin{figure}
\caption{The x-ray diffraction patterns (Cu K$_\alpha$) of the oxides, Ca$_{3-x}$Y$_x$LiRuO$_6$ (x= 0.0, 0.5 and 1.0)}
\end{figure}

\begin{figure}
\caption{(a) Magnetic suscptibility ($\chi$) as a function of temperature for Ca$_3$LiRuO$_6$ measured in the presence of a magnetic field of 2 kOe for ZFC and FC conditions of the specimen. (b) $\chi$ as a function of T for field-cooled state of Ca$_{3-x}$Y$_x$LiRuO$_6$ (x= 0.0, 0.5 and 1.0) oxides (H= 100 Oe). The lines drawn through the data points serve as guides to the eyes.} 
\end{figure}

\begin{figure}
\caption{The temperature dependence of inverse $\chi$ (H= 2 kOe)for the oxides Ca$_{3-x}$Y$_x$LiRuO$_6$ 
(x= 0.0, 0.5 and 1.0) in the paramagnetic state. A straight line is drawn through the linear region.} 
\end{figure}

\begin{figure}
\caption{Isothermal magnetization at selected temperatures for Ca$_{3-x}$Y$_x$LiRuO$_6$ (x= 0.0, 0.5 and 1.0) for the zero-field-cooled state of the specimens; typical low temperature hysteresis loops are shown for two compositions as insets. The lines drawn through the data points serve as guides to the eyes.}  
\end{figure}

\begin{figure}
\caption{ The electrical resistance ($\rho$) as a function of temperature for the oxides Ca$_{3-x}$Y$_x$LiRuO$_6$ (x= 0.0, 0.5 and 1.0).}
\end{figure}

\begin{figure}
\caption{The plot of $ln$$\rho$ versus inverse T for the oxides Ca$_{3-x}$Y$_x$LiRuO$_6$ (x= 0.0, 0.5 and 1.0.)}
\end{figure}
 
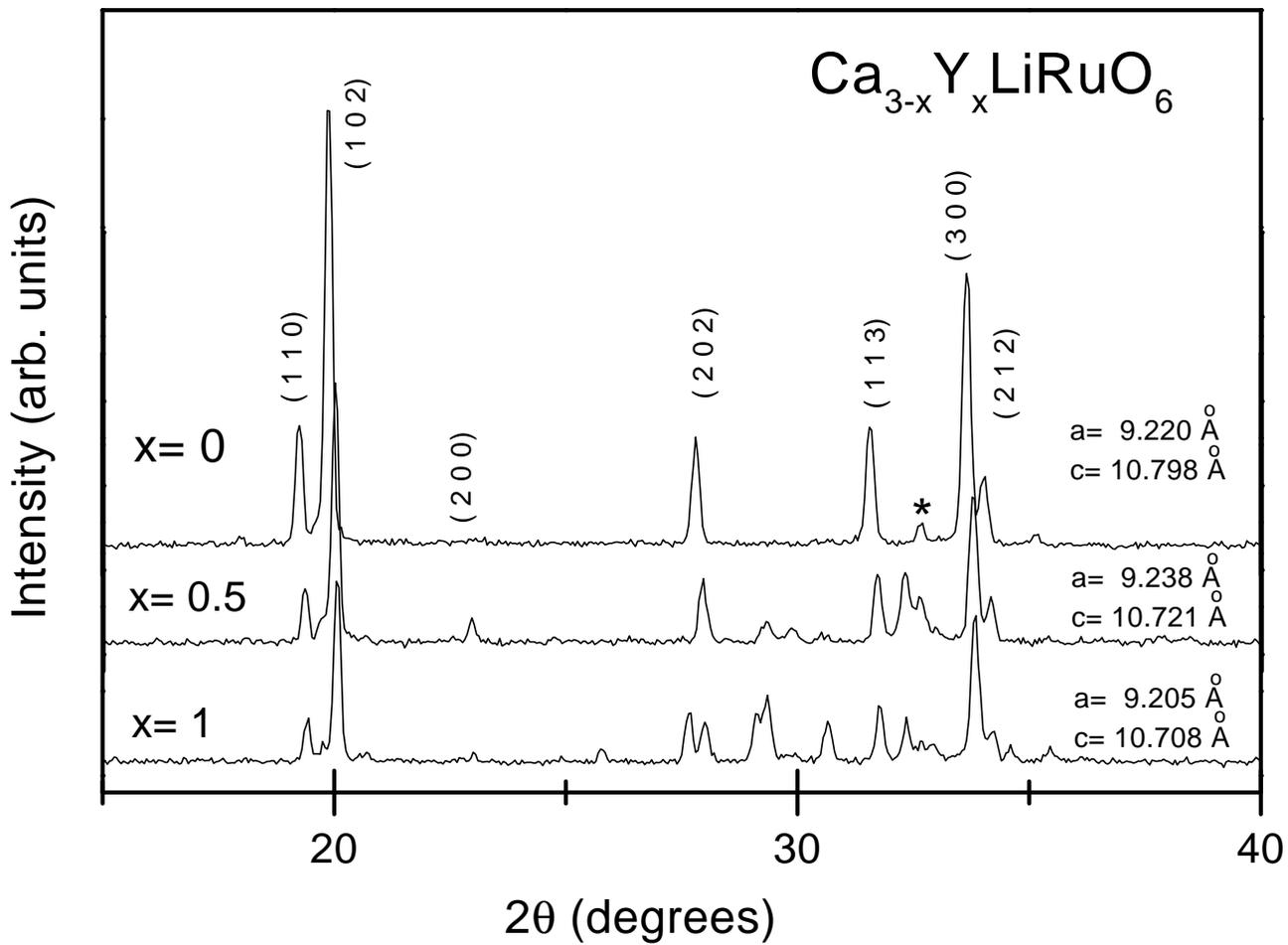
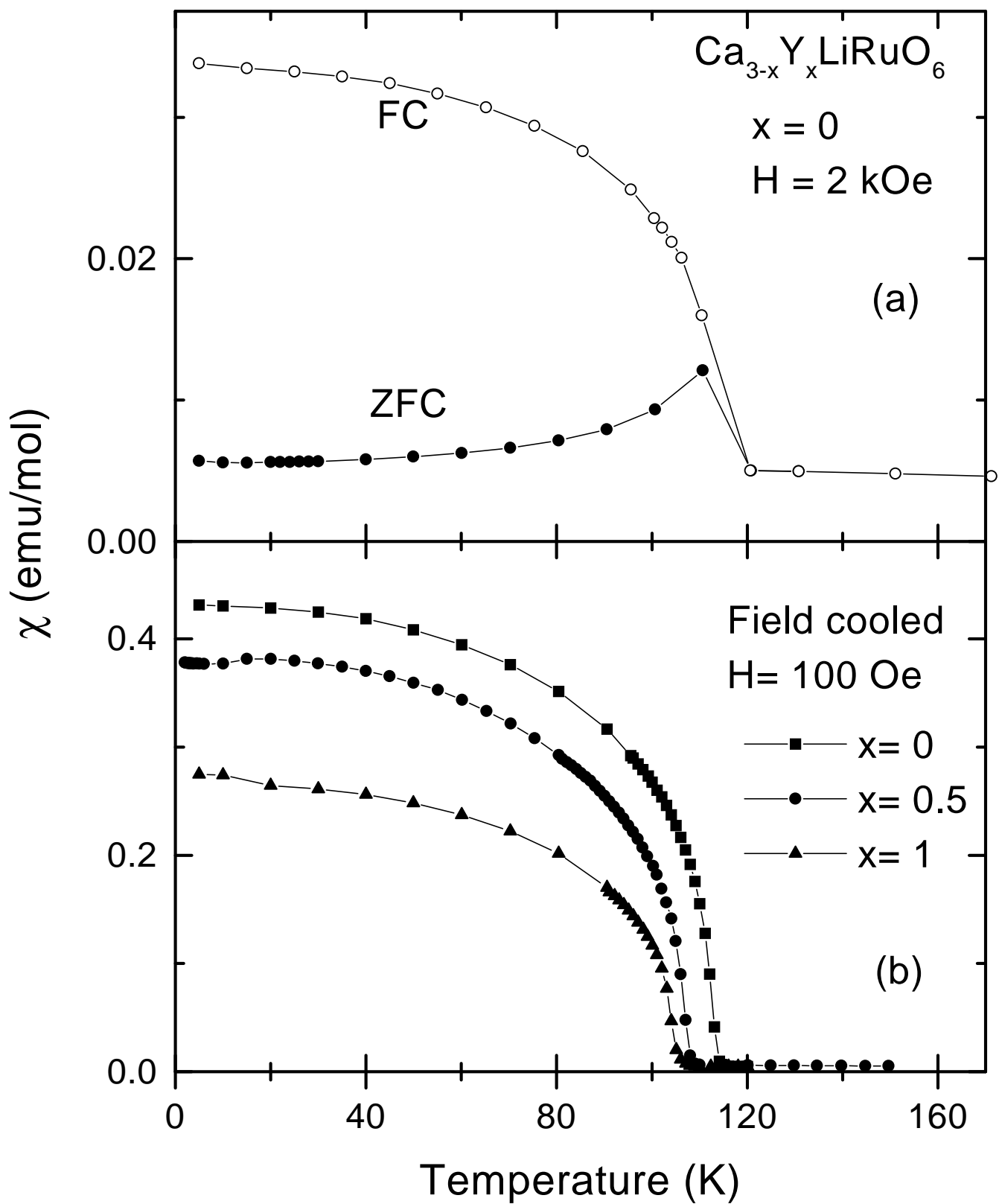
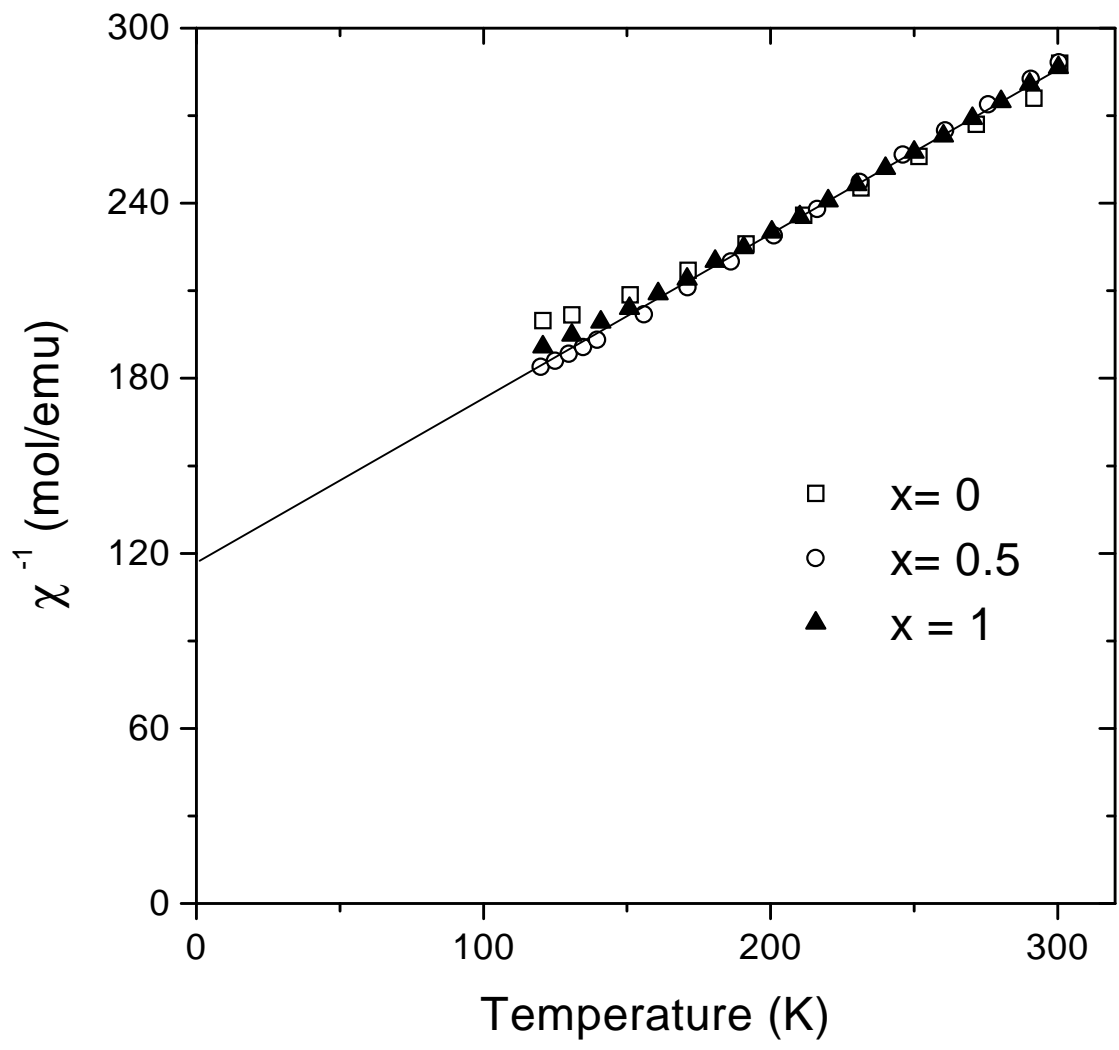
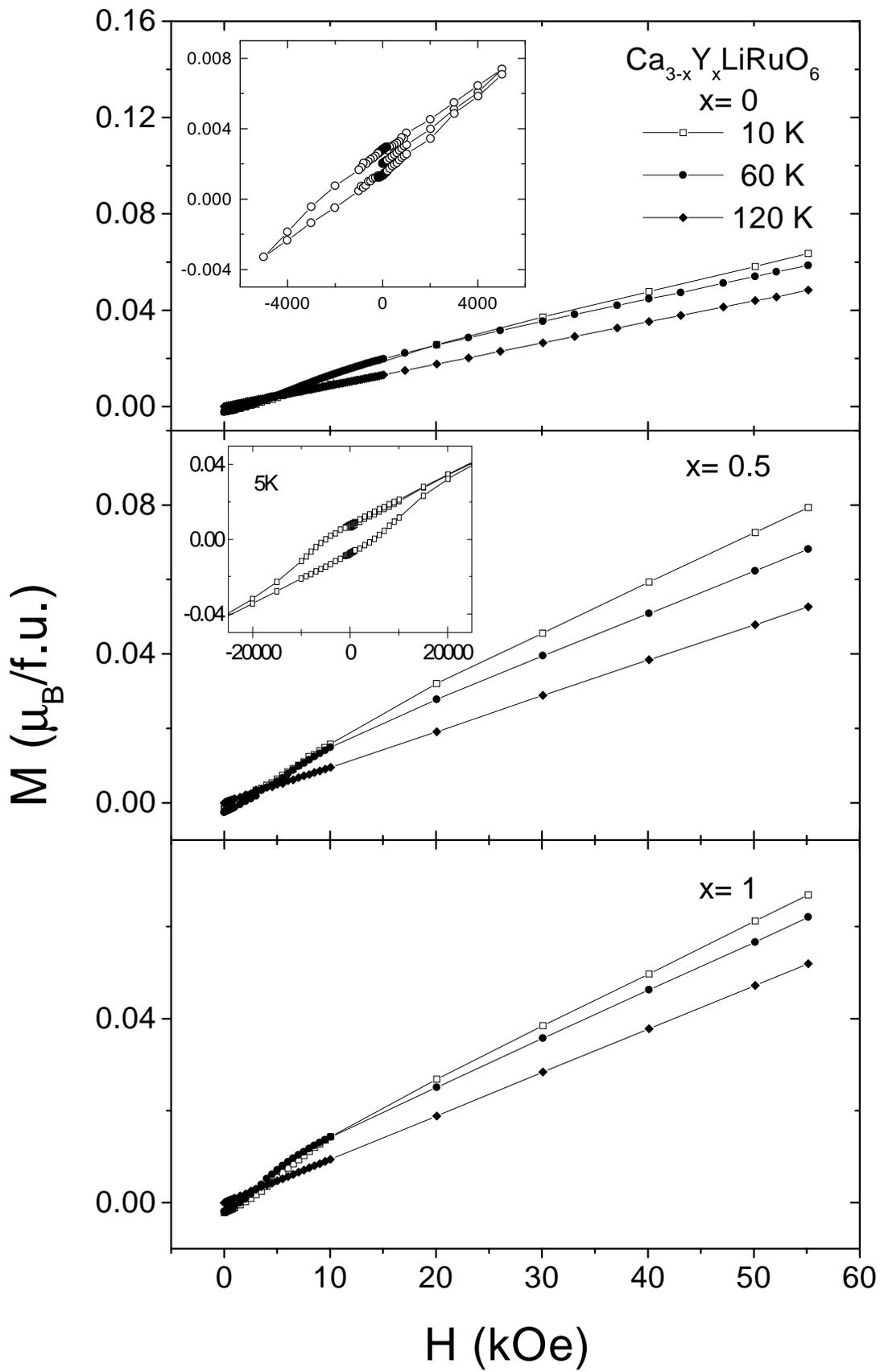
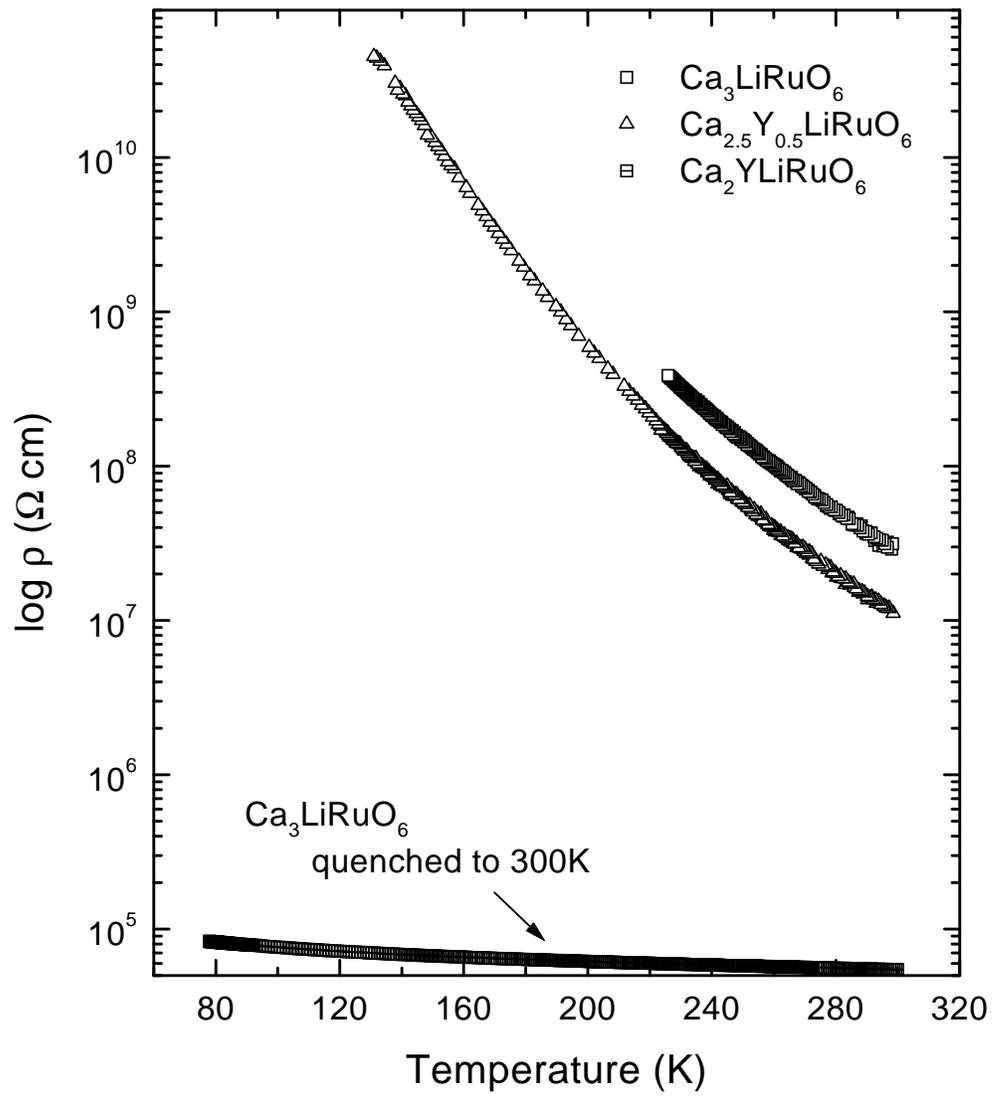
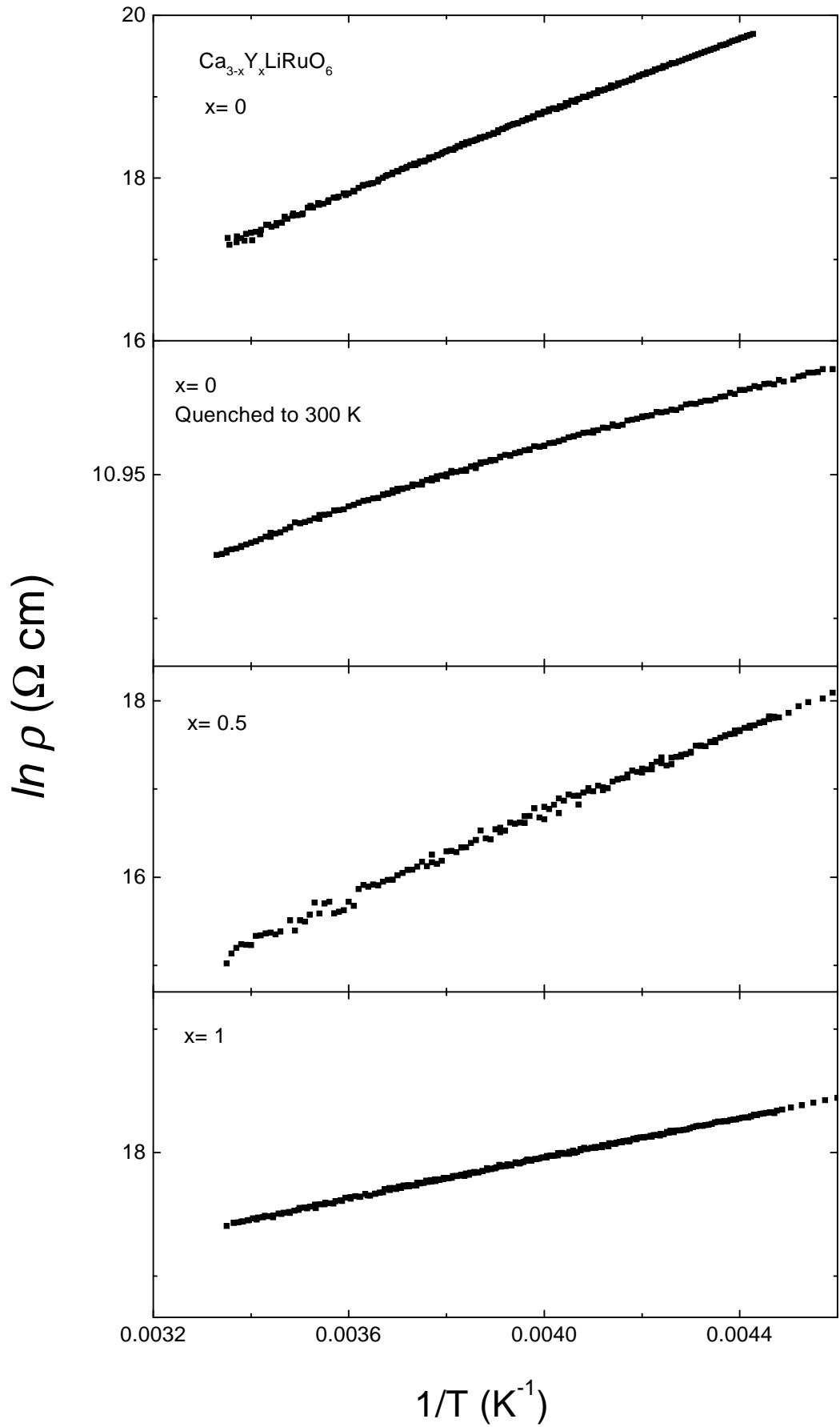
\end{document}